\documentclass[12pt,oneside]{article}
\usepackage[a4paper, left=2cm, right=2cm, top=2cm, bottom=2cm]{geometry}
\usepackage{graphicx} 
\usepackage{caption}
\usepackage{subcaption}
\usepackage{cite}
\usepackage{tabularx}
\usepackage{multirow}
\usepackage{authblk}  
\usepackage{enumitem}

\title{Limited Diffusion of Silicon in GaN: A DFT Study Supported by Experimental Evidence}
\author[1]{Karol Kawka}
\author[1]{Pawel Kempisty}
\author[2]{Akira Kusaba}
\author[1]{Krzysztof Golyga}
\author[1]{Karol Pozyczka}
\author[1]{Michal Fijalkowski}
\author[1]{Michal Bockowski}

\affil[1]{Institute of High Pressure Physics, Polish Academy of Sciences, Sokolowska 29/37, 01-142, Warsaw, Poland}
\affil[2]{Research Institute for Applied Mechanics, Kyushu University, Fukuoka 816-8580, Japan}

\begin{document}
	
\maketitle
	
\section{Introduction}

Gallium nitride (GaN) has emerged as a cornerstone material in modern optoelectronics and power electronics, widely utilized in applications such as light-emitting diodes (LEDs) and high-electron-mobility transistors (HEMTs). Its wide bandgap, high thermal conductivity, and excellent breakdown voltage make it superior to silicon-based technologies for high-frequency and high-power applications. The continuous advancements in GaN-based devices are driving the development of more efficient energy conversion systems, high-speed communication infrastructure, and next-generation semiconductor technology.

Silicon (Si) is the primary intentional donor used for fabricating \textit{n}-type GaN. In several epitaxial techniques, such as metal-organic vapor phase epitaxy (MOVPE) \cite{KOLESKE200255,Deatcher}, molecular beam epitaxy (MBE) \cite{Sawicka, Laukkanen}, and hydride vapor phase epitaxy (HVPE) \cite{Prozheev, Richter2013}. Si doping is achieved through its incorporation during the growth process. However, the in-situ doping method has its drawbacks, primarily the lack of precise control over the lateral spatial distribution of the doped regions. For some devices applications, precise local doping is required. In epitaxial growth techniques, this can be achieved through multi-step processes involving masking and material etching, making the fabrication process highly complex \cite{Langpoklakpam}.

Another approach for fabricating GaN-based transistors involves ion implantation, which enables precise doping control. However, this process disrupts the crystalline order, necessitating high-temperature annealing to restore the lattice structure and activate the dopants. Recent experimental studies have investigated the behavior of Mg and Be as an acceptor dopant in GaN, revealing its diffusion characteristics under different conditions by ultra-high-pressure annealing (UHPA)\cite{sierakowski, electronics9091380}. Similarly, the diffusion of Si in GaN should be carefully investigated.

Sharp Si profiles at interfaces between periodically doped and undoped material are commonly observed during high-temperature epitaxial growth of GaN \cite{kaess-jap-2016}. This behavior suggests that Si atoms exhibit very low mobility in GaN. Consequently, Si diffusion has not received much attention due to its challenging observability. However, some studies suggest that at temperatures significantly higher than typical epitaxial growth temperatures, diffusion can be activated. This aspect warrants further investigation, particularly in the context of the UHPA technique.

Jakiela et al. demonstrated that Si diffusion in epitaxial GaN occurs through two distinct mechanisms: rather slow bulk diffusion and significantly faster grain boundary/dislocation diffusion, with relatively low activation energies of 1.55 eV and 0.89 eV, respectively \cite{Jakiela2006}. Their study was conducted on Si-implanted GaN samples annealed in the temperature range of 900--1200 $^{\circ}$C. Other studies suggest that Si diffusion in AlN follows a concentration-dependent mechanism mediated by dopant-vacancy pairs, with an exceptionally high activation enthalpy of $10.34 \pm 0.32$ eV \cite{bonito_oliva}. However, in this case, the silicon source was an amorphous Si$_{1-x}$N$_x$ layer sputtered onto low-dislocation bulk AlN. Recent studies on GaN also indicate that Si diffusion at higher temperatures is negligible, contradicting earlier reports that suggested measurable diffusion in this range \cite{iwinska-spie-2022}.

To address these conflicting reports, we employ first-principles calculations to investigate Si diffusion in GaN. Although Si is the primary donor dopant in GaN, limited theoretical studies have explored its diffusion mechanisms. In this work, we compute the minimum energy paths (MEPs) for the vacancy-mediated diffusion mechanism of Si atoms in the Ga sublattice. The calculated energy barriers are then used to estimate the diffusion coefficient in various crystallographic directions. Theoretical considerations are compared with the results of UHPA experiments to elucidate observed discrepancies and improve the understanding of Si diffusion in GaN. These diffusion experiments are conducted at very high temperatures, exceeding 1100$^{\circ}$C, which are higher than those typically used in epitaxial growth experiments.

\section{Methods and models}

The presented research is based on Density Functional Theory (DFT) calculations within the Generalized Gradient Approximation (GGA). The exchange-correlation energy functional was parameterized using the Perdew-Burke-Ernzerhof (PBE) potential, incorporating the Jellium Surface (Js), Jellium Response (Jr), and Lieb-Oxford bound criteria \cite{PhysRevB.79.201106}. The SIESTA software (Spanish Initiative for Electronic Simulations with Thousands of Atoms) \cite{siesta, siesta2}, which employs the pseudopotential method and numerical atomic orbitals as basis sets, was used to determine the total energy, optimize the geometry, and perform phonon calculations for the investigated systems.

Pseudopotentials were generated using the following valence electron configurations: $2s^{2}$ and $2p^{3}$ for nitrogen, $3d^{10}$, $4s^{2}$, and $4p^{1}$ for gallium, and $3s^{2}$ and $3p^{2}$ for silicon. For all elements, triple-zeta (TZ) functions were used for the $s$ and $p$ orbitals, while double-zeta (DZ) functions were used for the $d$ orbitals. A supercell consisting of $3a\sqrt{3} \times 3a\sqrt{3} \times 3c$ GaN unit cells was employed as a bulk model, comprising a total of 324 atoms. Given the relatively large size of this system, only the $\Gamma$-point was used for Brillouin zone sampling. Next, two gallium atoms were removed from the supercell, and one silicon atom was substituted in place of Ga, resulting in a model containing one dopant atom and one vacancy.

The lattice constants obtained from the applied DFT parameterization were $a = 3.214$\r{A} and $c = 5.231$\r{A}, which remain in good agreement with the experimentally measured values for GaN: $a = 3.189$\r{A} and $c = 5.185$\r{A} \cite{leszcz-apl-1996}.

The calculations were carried out in the following steps:
\begin{enumerate}[label=\textit{(\roman*)}, noitemsep, topsep=0pt]
\item Relaxation of the geometry of the structure containing both point defects (Si$_\mathrm{Ga}$ and a vacancy);
\item Calculation of the minimum energy path (MEP) for the defect migration process using the Nudged Elastic Band (NEB) method;
\item Phonon calculations performed for both the relaxed system and the system corresponding to the saddle point on the MEP.
\end{enumerate}

Structural relaxation was carried out using the Fast Inertial Relaxation Engine (FIRE) algorithm \cite{Bitzek}. The structures were optimized until the force on each atom dropped below $10^{-3}$ eV/\r{A}. The convergence criterion SCF.DM.Tolerance for the self-consistent field (SCF) calculations was set to $10^{-4}$.

NEB calculations were performed using the Atomic Simulation Environment (ASE) software with SIESTA as the force and energy calculator \cite{ase-paper}. The ASE algorithms interpolate a path between the initial and final structures by generating intermediate configurations. Three crystallographic directions were considered in this study: a-direction [11$\bar{2}$0], c-direction [0001], and m-direction [1$\bar{1}$00]. Along the $a$ and $c$ directions, the path consisted of five structures in total: the initial structure before migration, the final structure after migration, and three intermediate structures. In the $m$ direction, due to the longer distance between the neighboring Si atom and the vacancy, the path included seven structures: the initial and final structures, along with five intermediate configurations. 

To search for the MEP, we interchangeably used two optimizers available in ASE: BFGS \cite{Packwood} and FIRE\cite{Bitzek}. A key parameter in NEB calculations is $f_{max}$, which represents the maximum force acting on any atom (in eV/\r{A}) and serves as a convergence criterion. In our calculations, $f_{max}$ was set to 0.01 eV/\r{A}. To accurately locate the saddle point along the MEP, the NEB method with the Climbing Image (CI-NEB) option was used. However, if convergence proved challenging, we employed a technique where the dopant atom was frozen very close to the expected saddle point while the rest of the structure was relaxed. This approach ensured that only a single negative phonon mode was present at the saddle point, which is necessary to confirm the correct transition state and compute reliable thermodynamic parameters. The energy differences between the system before and after this additional relaxation were negligible, on the order of $10^{-4} - 10^{-5}$ eV, while atomic forces were significantly reduced.

Afterwards, phonon calculations were performed within the harmonic approximation using the direct method. The force constant matrix was obtained by collecting forces acting on atoms displaced by 0.016~\r{A} from their equilibrium positions, one by one. Using the included \textit{Vibra} software, phonon frequencies were calculated in the wave-vector space with a $4 \times 4 \times 4$ mesh for the supercell. The thermodynamic parameters of the system, including zero-point vibrational energy, thermal energy, specific heat capacity, vibrational entropy, and vibrational free energy, were obtained from the phonon density of states.

The vibrational contribution to the free energy was determined using the following expression:
\begin{equation} \label{eq:free}
F^{vib}(T) = \frac{1}{2} \sum_{i,q} \hbar\omega_i(q) + k_{B}T \sum_{i,q} \ln\left[ 1-\exp\left(-\frac{\hbar\omega_i(q)}{k_BT}\right) \right]
\end{equation}
where $\omega_i$ are the phonon frequency, $\hbar$ is the reduced Planck constant, $k_{B}$ is the Boltzmann constant. The indices $i$ and $q$ denote the summation over all phonon modes and all sampling points in the wave vector space. The first term on the right-hand side is equivalent to the zero-point vibrational energy.

These parameters were used to determine the diffusion coefficient (D) using its microscopic definition. This expression was derived based on transition state theory, which describes diffusion as a hopping process between an initial and a final state, occurring through a saddle point on the potential energy surface.
\begin{equation}
\label{eq:dyf_full}
    D = f l^{2} \frac{k_{B}T}{h} \exp\Bigl(-\frac{\Delta E^{DFT}}{k_{B}T}\Bigr)\exp\Bigl(-\frac{\Delta F^{vib}(T)}{k_{B}T}\Bigr)
\end{equation}
where $\Delta E^{DFT}$ is the energy barrier obtained from DFT total energy calculations, and $\Delta F^{vib}(T)$ is the temperature-dependent difference in vibrational free energy between the equilibrium and saddle points. The parameter $l$ represents the jump distance, i.e., the displacement of the atom to a neighboring site, while $f$ is a correlation factor that accounts for the number of possible jump sites in a given lattice type and the effective diffusion dimensionality. For a hexagonal close-packed (\textit{hcp}) lattice, this factor is equal to 0.781 \cite{compaan}.

As mentioned, in the present study, we focus mainly on a vacancy-mediated mechanism by determining the minimum energy path of a silicon (Si) atom substituting a gallium (Ga) atom as it moves from its initial substitutional site to an adjacent vacancy site. Consequently, the effective diffusion coefficient should also account for the probability of finding a vacancy in the vicinity of the Si atom to which it can migrate. In the simplest approach, this probability is directly related to the concentration of such vacancies ($c_{V_{Ga}}$), which must therefore be considered.
\begin{equation}
\label{eq:dyf_eff}
D^{*} = c_{V_{Ga}} D
\end{equation}
In general, knowing the defect formation energy allows one to estimate its concentration. However, in practice, this is quite challenging due to the complex interactions between different types of defects and the necessity of accounting for various environmental conditions in which the process occurs. Moreover, the defect concentration does not necessarily have to correspond to the thermodynamically predicted equilibrium state, a metastable state may exist. It is well known that the concentration of point defects depends on the crystallographic plane on which the material was grown. Therefore, we do not delve into this issue and instead focus on aspects related to defect migration, treating the vacancy concentration as an independent parameter, which is of course a simplification. Moreover, we also examined several direct swap cases without any vacancies to verify whether this mechanism could serve as an alternative pathway for Si atom migration.

The diffusion coefficient D determined from Eq.~\ref{eq:dyf_full} can be related to its empirical counterpart, which is commonly fitted to experimental data:
\begin{equation}
\label{eq:diffusion}
  D = D_{0} \exp\Bigl(-\frac{\Delta H}{k_{B}T}\Bigr)
\end{equation}
where $\Delta H$ is the activation enthalpy of diffusion, and $D_{0}$ denotes the pre-exponential factor, also referred to as the frequency factor.

To relate our computational results more broadly to experimental conditions, we investigated both neutral and \textit{n}-type doping scenarios. The \textit{n}-type system was modeled by introducing three additional electrons into the supercell, simulating the formation of a triply negatively charged gallium vacancy (V$_{\mathrm{Ga}}^{3-}$). This condition is consistent with theoretical predictions indicating that gallium vacancies in GaN can accommodate up to three extra electrons, significantly lowering their formation energy under $n$-type conditions \cite{Lyons2017}. Previous studies have shown that donor-vacancy complexes, including silicon donors, exhibit charge states that shift the Fermi level within the bandgap, influencing defect stability and diffusion pathways \cite{baker_zlatko}. By incorporating this charge state into our simulations, we aim to provide a more accurate theoretical representation of the diffusion mechanisms observed experimentally in Si-doped GaN.

\section{Results and discussion}

\subsection{Electronic properties of relaxed systems}

\begin{figure}[t]
\centering
\includegraphics[]{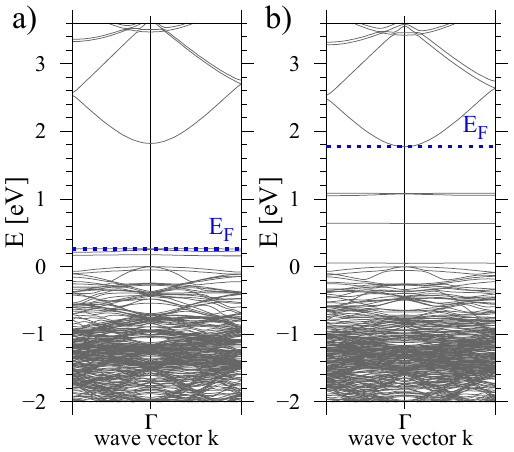}
\caption{Section of the band structure near the $\Gamma$ point for a GaN supercell containing both a substitutional Si$_{\mathrm{Ga}}$ defect and a gallium vacancy V$_{\mathrm{Ga}}$, with: a) neutral charge and b) negative charge (-3).}
\label{fig:bandstructure}
\end{figure}

Figure \ref{fig:bandstructure} shows the band structures of both the neutral and \textit{n}-type relaxed systems representing bulk GaN with a substitutional Si$_{\mathrm{Ga}}$ defect and a gallium vacancy, V$_{\mathrm{Ga}}$. In the neutral system, the presence of a Ga vacancy introduces deep triple acceptor states located above the valence band, where the Fermi level (E$_{\mathrm{F}}$) is pinned. In contrast, adding three electrons to this system shifts the Fermi level to the conduction band minimum (CBM).
However, no distinct state originating from Si dopants is observed near the CBM, which is a common issue in standard DFT-GGA calculations. Instead, this state is located deep in the conduction band. Nevertheless, the overall picture of the Fermi level shifting across the bandgap remains consistent with the notation defining the \textit{n}-type conductivity.

While analyzing the energies of the relaxed systems, significant energy differences were observed between configurations where the Si$_{\mathrm{Ga}}$ and V$_{\mathrm{Ga}}$ defects were either first or subsequent neighbors. Systems in which the vacancy is located at a distance of $\sqrt{3}a$ from the Si atom exhibit energies approximately 0.5~eV higher than those where the defects are separated by only one lattice constant $a$. These findings suggest the possibility of defect complex formation, as such configurations are energetically favorable.

\subsection{Diffusion energy barrier}

\begin{figure*}[b]
\centering
\includegraphics[]{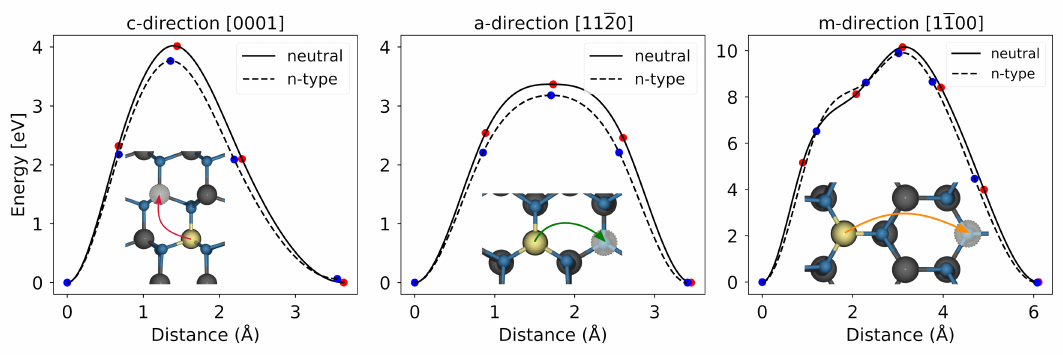}
\caption{Energy barriers for the diffusion of a Si atom in GaN via the vacancy-mediated mechanism along different crystallographic directions: $c$, $a$, and $m$.}
\label{fig:NEB_acm}
\end{figure*}

Figure~\ref{fig:NEB_acm} presents a summary of the energy barriers for the diffusion of a Si atom in GaN via the vacancy-mediated mechanism along different crystallographic directions. The illustrative geometric configurations of the lattice sites are shown in the insets. All calculated barriers are relatively high. The lowest one, corresponding to the lateral movement of a Si atom to its nearest neighbor along the $a$-direction, is approximately 3.2 eV, as shown in Fig.~\ref{fig:NEB_acm}b. For hopping along the $c$-direction, the barrier increases to 3.8 eV.

Notably, these lower values were obtained for migration in an electron-rich system. In neutral systems, the barriers are slightly higher, by approximately 0.2 eV and 0.25 eV, respectively. Exceptionally high energy barriers, reaching up to 10 eV, were obtained for migration along the $m$-direction to the next distant neighbor. Therefore, this direct migration scenario is highly unlikely. Instead, one should expect that the migration to the target site located at a distance of $\sqrt{3}a$ will more likely occur as a sequence of two shorter hops, each with a lower energy barrier.

\begin{figure*}[t]
	\centering
	\includegraphics[width=0.97\textwidth]{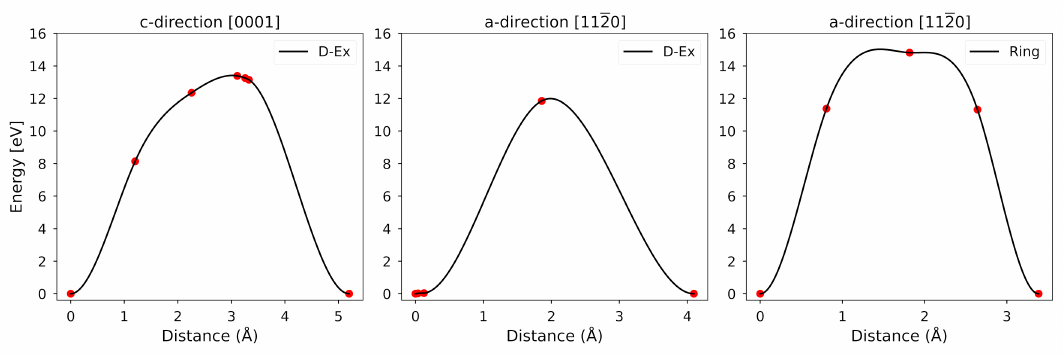}
	\caption{Energy barriers for various diffusion mechanisms. The left graph presents the minimum energy path (MEP) for the direct exchange (D-Ex) mechanism along the $c$-direction, the center graph depicts the MEP for the same mechanism along the $a$-direction, and the right graph illustrates the MEP for the ring mechanism along the $a$-direction.}
	\label{fig:big_barriers}
\end{figure*}

Similarly, migration scenarios that do not involve vacancies are highly unlikely. The exchange of atoms can occur either directly between two neighboring Si and Ga lattice sites via the direct exchange (D-Ex) mechanism or through a ring-like motion. The calculated energy barriers for such processes, both along the $a$- and $c$-directions, are very high -- on the order of 12 eV or higher, as shown in Fig.~\ref{fig:big_barriers}. All calculations were performed for neutral systems.

The first two cases correspond to the D-Ex mechanism between two adjacent atoms: Si$_{\mathrm{Ga}}$ and Ga, along the [0001] and [11$\bar{2}$0] directions, with energy barriers of 13.4 eV and 11.8 eV, respectively. The last case involves a ring-like motion of three atoms: Si$_{\mathrm{Ga}}$ and two Ga atoms. In this case, the energy barrier was found to be 14.8 eV.

\begin{figure}[h]
	\centering
	\includegraphics[]{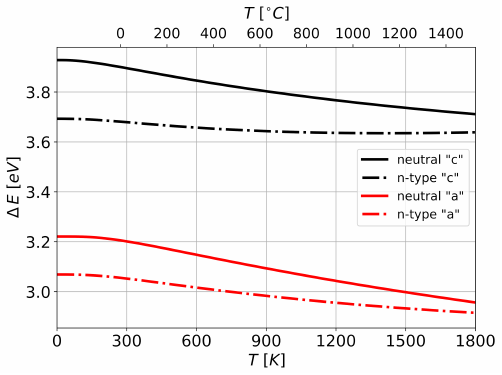}
	\caption{Change of the effective energy barrier to vacancy-mediated migration of Si atom in GaN as a function of temperature.}
	\label{fig:f_vib}
\end{figure}

It should be noted that the energy barriers determined based on the total energy of the systems effectively correspond to very low-temperature conditions. To accurately describe diffusion processes at elevated temperatures, it is essential to consider changes in free energy, primarily accounting for vibrational degrees of freedom. Figure~\ref{fig:f_vib} presents the variation of the effective energy barrier as a function of temperature after incorporating the contribution of vibrational free energy ($\mathrm{F}^{vib}$) calculated from the phonon spectra according to Eq.~\ref{eq:free}. As observed, in neutral systems, the effective barrier systematically decreases with increasing temperature. However, this effect is relatively weak, with a total reduction of approximately 0.25 eV between the temperatures of $0$ K and $1800$ K. Furthermore, in $n$-type doped systems, the reduction in the energy barrier is significantly smaller; for diffusion along the $c$-direction, it is almost negligible, and the barrier can be considered effectively constant. Overall, it should be emphasized that the absolute values of these barriers remain relatively high, meaning that diffusion will still be limited even at elevated temperatures.

\subsection{Diffusion coefficient}

Based on Eq.~\ref{eq:dyf_full}, the values of the silicon diffusion coefficient in GaN along selected crystallographic directions can be determined as a function of temperature. Here, we focus on presenting its values for the vacancy-mediated mechanism in the $a$- and $c$-directions.

\begin{figure}[b] 
\centering \includegraphics[]{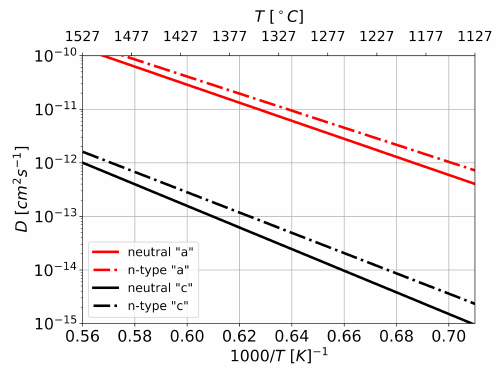} 
\caption{Diffusion coefficient of Si in GaN along the $a$-direction (red) and $c$-direction (black).} 
\label{fig:coefficient} 
\end{figure}

To connect theoretical predictions with experimental observations, the diffusion coefficient $D$ is typically plotted on a logarithmic scale as a function of $1/T$, as shown in Fig.~\ref{fig:coefficient}. This is a standard procedure for determining the activation enthalpy and the pre-exponential factor $D_{0}$ by fitting to the linearized form of the Arrhenius equation:
\begin{equation} 
\label{lnD} 
\ln D = -\frac{\Delta H}{k_{B}}\times\frac{1}{T} + \ln{D_{0}} 
\end{equation}
Notably, the diffusion coefficient in the lateral direction is expected to be approximately two orders of magnitude higher than in the polar direction. Even at extremely high temperatures, around 1800 K, these values do not exceed $10^{-9}$ cm$^{2}$/s, which indicates a relatively low diffusion rate. However, at this stage, Figure~\ref{fig:coefficient} does not account for the factor related to vacancy concentration; therefore, the values presented represent the maximum possible limits. In reality, the observed values will be lower, proportional to the vacancy concentration, and are typically expected to be at least three orders of magnitude lower, as the V$_{\mathrm{Ga}}$ concentration does not exceed $10^{18}$ cm$^{-3}$.

It should be noted that these values are provided under the assumption that the conditions ensure the thermodynamic stability of GaN. Considering that increased pressure is required to maintain these conditions, additional calculations should be performed to account for pressure-induced changes in the GaN lattice constants. However, this effect is not expected to significantly alter the diffusion coefficient and may even lead to its reduction.

Table~\ref{tab:diffusion_results} presents a summary of the results obtained from NEB calculations, along with the pre-exponential diffusion coefficient $D_{0}$ and activation enthalpy, which were derived through a form of reverse engineering based on the diffusion coefficient determined from the microscopic theory of diffusion (Eq.~\ref{eq:dyf_full}). As can be seen, the energy barrier values obtained from the NEB calculations and the fitted activation enthalpy are nearly identical. This outcome is not entirely obvious, given that in some cases, we observed a temperature dependence of the effective barrier, and the pre-exponential factor exhibits a strong dependence on temperature. For the neutral system at the saddle point along the $m$-direction, obtaining a correct phonon dispersion within the harmonic approximation was not possible due to the presence of two negative phonon modes. This indicates that the saddle point was not determined with sufficient precision. Nevertheless, the contribution from $\Delta F^{vib}$ has a negligible impact on the energy barrier height, given its high absolute value.

\begin{table}[b]
	\caption{Comparison of characteristic diffusion parameters obtained from DFT calculations and the fitting method to the Arrhenius equation.}
	\label{tab:diffusion_results}
	\centering
	\renewcommand{\arraystretch}{1}  
	
	\begin{tabular}{ |c|>{\centering\arraybackslash}p{4cm}|>{\centering\arraybackslash}p{4cm}|>{\centering\arraybackslash}p{4cm}| }
		\hline
		\textbf{Case} & \textbf{NEB energy barrier [eV]} & \textbf{Activation enthalpy [eV]} & \textbf{$D_{0}$ coefficient [$cm^{2}/s$]} \\
		\hline
		\multicolumn{4}{|c|}{a-direction [11$\bar{2}$0]} \\
		\hline
		neutral  & 3.366  & 3.344  & 0.373 \\ 
		n-type   & 3.179  & 3.163  & 0.151 \\
		\hline
		\multicolumn{4}{|c|}{c-direction [0001]} \\
		\hline
		neutral  & 4.014  & 4.005  & 0.203 \\
		n-type   & 3.766  & 3.758  & 0.065 \\
		\hline
		\multicolumn{4}{|c|}{m-direction [1$\bar{1}$00]} \\
		\hline
		neutral  & 10.148  & --  & -- \\
		n-type   & 9.887   & 9.898  & 2.017 \\
		\hline
	\end{tabular}
	
\end{table}

\subsection{UHPA annealing of Si-implanted GaN samples}

\begin{figure}[b]
	\centering
	\includegraphics{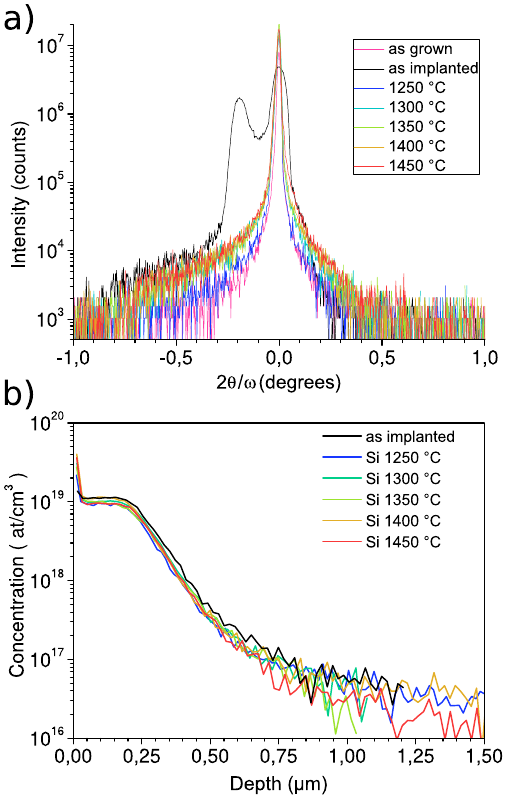}
	\caption{Results of characterization of the MOVPE-GaN/Ammono-GaN sample (10-${\mu}$m-thick, TDD = 5 $\times10^4$ cm$^{-2}$), implanted with Si (energy: 200 keV, dose: 3$\times$$10^{14}$ cm$^{-2}$) and annealed under UHPA conditions (1 GPa, 30 minutes): (a) 2$\theta$/$\omega$ XRD scan; (b) SIMS depth profile of Si.}
	\label{fig:xrd}
\end{figure}

\begin{figure}[b]
	\centering
	\includegraphics{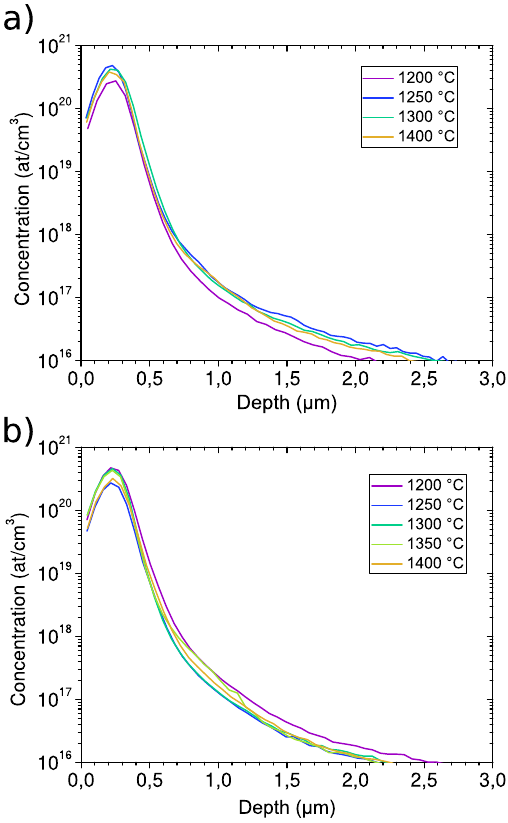}
	\caption{SIMS-measured Si concentration profiles in HVPE-GaN after UHPA processing (3 hours at 1 GPa): (a) 20-$\mu$m-thick HVPE-GaN/Ammono-GaN samples with a threading dislocation density of 5 $\times10^4$ cm$^{-2}$; (b) 20-$\mu$m-thick HVPE-GaN/sapphire samples with a threading dislocation density of $10^8$ cm$^{-2}$.  Si was implanted with a beam energy of 300 keV and a dose of 1$\times$$10^{15}$ cm$^{-2}$.}
	\label{fig:sims}
\end{figure}

To investigate the diffusion behavior of Si in GaN, experiments were conducted on ion-implanted GaN samples subjected to UHPA. XRD (X-Ray Diffraction)  and SIMS (Secondary Ion Mass Spectrometry) measurements were performed to evaluate both implantation damage recovery and potential Si diffusion, providing experimental insight into the stability of Si profiles in GaN.
Figure \ref{fig:xrd}a demonstrates a XRD scan of an 10-${\mu}$m-thick MOVPE-GaN sample grown on Ammono-GaN substrate, with a threading dislocation density (TDD) of 5 $\times10^4$ cm$^{-2}$, implanted with Si with an energy of 200 keV and a dose of 3$\times$$10^{14}$ cm$^{-2}$. The damage caused by the ion implantation process is clearly visible as a broadened signal with two distinct maxima. The other XRD signals demonstrate that UHPA at temperatures above 1300\,$^\circ$C under 1\,GPa of N$_2$ pressure for 30 minutes effectively removes implantation-induced damage and restores the high-quality crystalline structure of the sample. Figure~\ref{fig:xrd}b shows SIMS-measured Si concentration profiles in GaN samples, taken immediately after implantation and following annealing. 
The minimal differences in the SIMS profiles indicate that Si diffusion is negligible, even under UHPA conditions.

For comparison, the procedure was repeated using 20-$\mu$m-thick HVPE-GaN samples, grown on Ammon-GaN and sapphire substrates, with an increased Si implantation energy and dose, up to 300 keV and 1$\times$$10^{15}$ cm$^{-2}$, respectively. The Si concentration profiles after UHPA annealing (3 hours, 1 GPa) at various temperatures also remain unchanged compared to the as-implanted profiles (see Fig.~\ref{fig:sims}a and Fig.~\ref{fig:sims}b), confirming that Si diffusion is negligible under these conditions. The use of different substrates resulted in varying threading dislocation densities, but this also did not affect silicon mobility.

\section{Summary}

In this study, the diffusion behavior of Si in GaN was investigated using first-principles DFT calculations and experimental validation through ultra-high-pressure annealing (UHPA). The results demonstrate that Si diffusion in GaN is significantly limited under typical processing conditions. Vacancy-mediated diffusion was identified as the primary mechanism, with energy barriers ranging from 3.2 eV along the a-direction to 3.8 eV along the c-direction, reflecting the intrinsic anisotropy of diffusion in GaN. Furthermore, direct migration along the m-direction was found to be highly improbable due to an exceptionally high energy barrier exceeding 10 eV. In such cases, diffusion is more likely to occur as a sequence of two shorter hops in a-directions, each with a lower energy barrier, rather than a single long-range migration event.
Additionally, alternative diffusion mechanisms, such as direct exchange and ring-like migration, were considered but deemed unlikely to occur due to the extremely high energy barriers, exceeding 12 eV, associated with these pathways.
The diffusion coefficients calculated from transition state theory confirmed extremely low diffusivity, with values not exceeding 10$^{-9}$ cm$^{2}$/s even at elevated temperatures up to 1800 K. However, these values represent the upper limit of diffusion, as they do not account for the concentration of defects, such as Ga vacancies, which would further reduce the effective diffusion rate. Experimental results from UHPA-treated Si-implanted GaN samples, including XRD and SIMS measurements, showed no detectable Si diffusion, further validating the theoretical predictions. Neither the use of GaN grown by a different method (MOVPE vs.\ HVPE), different substrates (Ammono-GaN vs.\ sapphire), varying threading dislocation densities (on the order of $10^4$–$10^8$\,cm$^{-2}$), nor changes in the implanted ion dose and energy led to increased Si diffusion. Changes in annealing conditions, such as temperature and time, also did not affect Si mobility.
These findings suggest that Si doping in GaN remains stable under standard device fabrication conditions, making it suitable for applications requiring precise n-type doping without significant diffusion-related broadening of doped regions. The study not only resolves previous inconsistencies in diffusion reports but also provides a comprehensive understanding of Si behavior in GaN, essential for advancing GaN-based electronic and optoelectronic devices.

\subsection*{CRediT authorship contribution statement}
Karol Kawka: Writing – original draft, Visualization, Investigation, Conceptualization, Formal analysis, Methodology; Pawel Kempisty: Supervision, Writing – review and editing, Funding acquisition, Conceptualization; Akira Kusaba: Validation, Writing – review and editing,; Krzysztof Golyga: Investigation, Visualization; Karol Pozyczka: Investigation; Michal Fijalkowski: Investigation, Visualization; Michal Bockowski: Funding acquisition,  Writing – review and editing.

\subsection*{Acknowledgments}
This research was supported by the National Science Centre of Poland, under grant numbers 2021/41/B/ST5/02764 and 2023/49/B/ST5/03319. This work was partly supported by the Collaborative Research Program of the Research Institute for Applied Mechanics, Kyushu University. We gratefully acknowledge Polish high-performance computing infrastructure PLGrid (HPC Center: ACK Cyfronet AGH) for providing computer facilities and support within computational grant no. PLG/2023/016827.
The authors thank prof. Filip Tuomisto of Helsinki University for Si implantation into GaN and prof. Rafal Jakiela for performing SIMS measurements.

\bibliographystyle{elsarticle-num} 

\end{document}